\documentclass[conference,10pt]{IEEEtran}
\usepackage{amsmath,graphicx,cite}
\usepackage[ruled,linesnumbered]{algorithm2e}
\usepackage{amsfonts}
\usepackage{multirow}
\usepackage{balance}

\IEEEoverridecommandlockouts

\newcommand{\A}{{\bf A}}

\newcommand{\x}{{\bf x}}

\newcommand{\I}{{\bf I}}

\newcommand{\wCxkt}{\widehat{\bf C}^{k,t}}

\begin{document}

\title{Algorithm for Independent Vector Extraction Based on Semi-Time-Variant Mixing Model}

\author{\authorblockN{Zbyn\v{e}k Koldovsk\'{y}\authorrefmark{1}, V\'aclav Kautsk\'y\authorrefmark{2}, Tom\'a\v{s} Kounovsk\'y\authorrefmark{1} and Jaroslav \v{C}mejla\authorrefmark{1}}
\authorblockA{\authorrefmark{1}Acoustic Signal Analysis and Processing Group, Faculty of Mechatronics, Informatics, 
and Interdisciplinary Studies,\\ Technical University of Liberec, Czech Republic.}
\authorblockA{\authorrefmark{2}Faculty of Nuclear Sciences and Physical 
Engineering, Czech Technical University in Prague, Czech Republic}
\thanks{This work was supported by 
		The Czech Science Foundation through 
		Project No.~20-17720S, and by the department of the Navy, Office of Naval Research Global, through Project No.~N62909-19-1-2105}
}

\maketitle
\begin{abstract}
A new algorithm for dynamic independent vector extraction is proposed. It is based on the mixing model where mixing parameters related to the source-of-interest (SOI) are time-variant while the separating parameters are time-invariant. A contrast function based on the quasi-likelihood approach is optimized using the Newton-Raphson approach. The update is computed without imposing the orthogonal constraint, and the orthogonality is enforced afterward. This yields an algorithm that is significantly faster than gradient-based algorithms while different from fixed-point methods, which are even faster. We show advantageous properties of the proposed algorithm compared to the fixed-point methods in an on-line processing regime where stable convergence to the SOI is the important issue. The effectiveness of the method is demonstrated in a speech extraction experiment with a dense microphone array.
\end{abstract}
\section{Introduction}
\label{sec:intro}
Independent Component and Vector Analysis (ICA/IVA) are popular methods for Blind Source Separation (BSS). The goal is to separate $K$ linear instantaneous mixtures of original signals into {\em independent} components that correspond to these signals up to indeterminable scales and order; $K=1$ in case of ICA. In IVA, $K>1$ mixtures are separated jointly based on a joint probabilistic source model where dependencies between the components that correspond to the same source are taken into account; these components form so-called vector component \cite{comon2010handbook,kim2007}. In Independent Component/Vector Extraction (ICE/IVE), the goal is to extract only a particular source-of-interest (SOI), i.e., to separate it from the other signals \cite{huber1985,koldovsky2019TSP,scheibler2019overiva,ikeshita2021independent}. 

Conventional methods assume the static linear mixing model given by
\begin{equation}\label{eq:modelstaticICA}
\x_k = \A_k {\bf s}_k,\qquad k=1,\dots,K,
\end{equation} 
where $\x_k$ is a $d\times 1$ vector of random variables (RVs) representing the mixed signals; $\A_k$ is a 
$d\times d$ non-singular time-invariant mixing matrix; and ${\bf s}_k$ is a $d\times 1$ vector of independent RVs representing the original signals; $k$ is the mixture index. The determined case, i.e. when $\A_k$ are square, is often preferred because then the goal can be formulated as to estimate square de-mixing matrices ${\bf W}_k$ such that ${\bf y}_k={\bf W}_k\x_k$ correspond to ${\bf s}_k$ up to the indeterminable scales and order. ${\bf W}_k$ are sought such that the signals ${\bf y}_k$ are as independent as possible \cite{comon1994}.

In many real-world situations, the sources are moving, which can be modeled by assuming that $\A_k$, hence also ${\bf W}_k$, are time-varying. Typically, adaptive methods are derived from off-line algorithms in a sequential manner so that $\A_k$ or ${\bf W}_k$ are progressively updated; see, e.g., \cite{taniguchi2014,khan2015,hsu2016}. Inherently, these methods apply the static mixing model \eqref{eq:modelstaticICA} on short intervals of data.
Recently, a novel dynamic mixing model for ICE/IVE has been introduced referred to as CSV (Constant Separating Vector) \cite{kautsky2020CRLB,jansky2020,koldovsky2020fastdiva}. In CSV, the mixing parameters related to the SOI are time-variant while the separating parameters are time-invariant. The idea behind is that an invariant separating vector extracts the SOI from the entire area of its motion (during the interval of data), which is delimited by the time-varying mixing parameters\footnote{Similar idea based on beam-widening appears in the literature on robust beamforming; see, e.g., \cite{sharon_book,li2006robust,vorobyov2004adaptive}.}. Advantageous theoretical properties and practical usefulness of CSV for off-line as well as on-line blind extraction of moving sources have been recently reported in \cite{kautsky2020CRLB,jansky2020,amor2021}.

In this paper, a new algorithm for IVE based on the CSV model, hereafter referred to as QuickIVE\footnote{Available at {\tt https://asap.ite.tul.cz/downloads/ice/}}, is proposed. It seeks optimum points of a quasi-likelihood function using the Newton-Raphson optimization. The update is computed without imposing the orthogonal constraint as compared to the FastDIVA algorithm in \cite{koldovsky2020fastdiva}, while the orthogonality is enforced afterwards. In experiments, the algorithm is considerably faster than gradient-based algorithms, comparable or faster than auxiliary function-based methods \cite{ono2011stable,scheibler2019overiva,jansky2020} while slower than fixed-point methods such as FastICA \cite{hyvarinen1999}, FastIVA \cite{lee2007fast} and FastDIVA \cite{koldovsky2020fastdiva}. Nevertheless, QuickIVE appears to be more useful in on-line processing deployments where fixed-point methods show high sensitivity to a varying non-Gaussianity of the SOI, which can cause their sudden divergence (in particular, convergence to a different source than the SOI).


The paper is organized as follows. The new algorithm is derived in Section~II and is compared with the fixed-point algorithms in Section~III. Section~IV is focused on numerical and experimental evaluations and comparisons with the other algorithms. Section~V concludes this paper. 

{\em Notations:} 
Plain, bold, and bold capital letters denote, respectively, scalars, vectors, and matrices. Upper index $\cdot^T$, $\cdot^H$, or 
$\cdot^*$ denotes, respectively, transposition, conjugate transpose, or complex 
conjugate. The Matlab convention for matrix/vector concatenation will be used, e.g., $[1;\,{\bf g}]=[1,\, {\bf g}^T]^T$. 
${\rm E}[\cdot]$ stands for the expectation operator, and $\hat{\rm E}[\cdot]$ is the sample-based average taken over all available samples of the argument; $\hat{a}$ denotes an estimate of $a$; $\{{\bf w}_k\}_k$ is a short notation of ${\bf w}_1,\dots,{\bf w}_K$; $\left< a_t\right>_t$ is a short notation of $\frac{1}{T}\sum_{t=1}^T a_t$.

\section{Proposed Algorithm}
\subsection{Problem formulation}
We will consider complex-valued signals and parameters, however, the conclusions are valid also for the real-valued case. Let $N$ samples of signals be available in $K$ mixtures, and let the samples be divided into $T\geq 1$ time-intervals called blocks, for simplicity, of the same length $N_b$; hence $N=T\cdot N_b$. CSV is a semi-time-varying model described by
\begin{equation}\label{eq:mixingmodel}
    {\bf x}_{k,t}={\bf A}_{k,t}{\bf s}_{k,t},
\end{equation}
where $k=1,\dots,K$ is the mixture index; $t=1,\dots,T$ is the block index; ${\bf A}_{k,t}$ is the square non-singular mixing matrix; ${\bf s}_{k,t}$ and ${\bf x}_{k,t}$ are RVs representing samples of the original and observed signals, respectively; ${\bf s}_{k,t}$ is divided as ${\bf s}_{k,t}=[s_{k,t};{\bf z}_{k,t}]$ where $s_{k,t}$ stands for the SOI while ${\bf z}_{k,t}$ represents {\em background}, which is an unspecified mixture of the other signals. The mixing matrices ${\bf A}_{k,t}$ are parameterized as
\begin{equation}\label{eq:CSVmixingmatrix}
 {\bf A}_{k,t} = 
\begin{pmatrix}
{\bf a}_{k,t} & {\bf Q}_{k,t}
\end{pmatrix}  = 
 \begin{pmatrix}
  \gamma_{k,t} & {\bf h}_k^H\\
   {\bf g}_{k,t} &   \frac{1}{\gamma_{k,t}}({\bf g}_{k,t}{\bf h}_{k}^H-\I_{d-1}) \\
    \end{pmatrix},   
\end{equation}
where ${\bf a}_{k,t}=[\gamma_{k,t};{\bf g}_{k,t}]$ is the {\em mixing vector} and ${\bf w}_{k}=[\beta_{k};{\bf h}_{k}]$ is the {\em separating} vector corresponding to the SOI in the $k$th mixture and $t$th block. Note that ${\bf a}_{k,t}$ depends on $t$ while ${\bf w}_{k}$ is constant over $t$. For $T=1$, CSV coincides with the time-invariant (static) IVE model \cite{koldovsky2019TSP}.

The vectors ${\bf a}_{k,t}$ and ${\bf w}_{k}$ are free vector parameters linked through ${\bf a}_{k,t}^H{\bf w}_{k}=1$ for all $t$ and $k$, which is referred to as the {\em distortionless constraint}. This constraint, together with the structure of ${\bf A}_{k,t}$ given by \eqref{eq:CSVmixingmatrix}, ensures that ${\bf A}_{k,t}^{-1}$ is equal to
\begin{equation}
 {\bf W}_{k,t} = 
 \begin{pmatrix}
 {\bf w}_{k}^H\\
 {\bf B}_{k,t}
 \end{pmatrix}  = 
 \begin{pmatrix}
 \beta_k^* & {\bf h}_k^H\\
  {\bf g}_{k,t} & -\gamma_{k,t}  \I_{d-1} \\
 \end{pmatrix},
\end{equation}
where $\I_{d}$ is the $d\times d$ identity matrix. It means that ${\bf W}_{k,t}$ separates ${\bf x}_{k,t}$ into $s_{k,t}={\bf w}_{k}^H\x_{k,t}$ and ${\bf z}_{k,t}={\bf B}_{k,t}\x_{k,t}$; see also Section~II in \cite{koldovsky2020fastdiva}.

\subsection{Contrast function}
Let $p({\bf s}_{t})$ and $p_{{\bf z}_{k,t}}({\bf z}_{k,t})$ denote the joint pdf of the SOI vector component ${\bf s}_t=[s_{1,t},\dots,s_{K,t}]^T$ and of ${\bf z}_{k,t}$, respectively. For simplicity, $p(\cdot)$ will be assumed constant over $t$, and ${\bf z}_{1,t},\dots,{\bf z}_{K,t}$ are assumed to be independent.

Since $p(\cdot)$ and $p_{{\bf z}_{k,t}}(\cdot)$ are not known in the blind setting, they have to be replaced by suitable model densities. In  \cite{koldovsky2020fastdiva} (Section~III.B), it justified that $p(\cdot)$ can be replaced by
\begin{equation}\label{eq:modeldensity}
    p({\bf s}_t) \approx f\left(\left\{\frac{s_{k,t}}{\hat\sigma_{k,t}}\right\}_k\right)\left(\prod_{k=1}^K\hat\sigma_{k,t}\right)^{-2},
\end{equation}
where $f(\cdot)$ should be a suitable normalized non-Gaussian pdf, and $\hat\sigma_{k,t}^2$ is the sample-based variance of the estimate of $s_{k,t}$; it holds that $\hat\sigma_{k,t}=\sqrt{{\bf w}_k^H\widehat{\bf C}^{k,t}{\bf w}_k}$ where $\widehat{\bf C}^{k,t}=\hat{\rm E}[{\bf x}_{k,t}{\bf x}_{k,t}^H]$ is the sample-based covariance matrix of ${\bf x}_{k,t}$.

The unknown $p_{{\bf z}_{k,t}}({\bf z}_{k,t})$ can be replaced by the zero mean circular Gaussian pdf $\mathcal{CN}(0,{\bf C}_{\bf z}^{k,t})$, where
${\bf C}_{\bf z}^{k,t}={\rm E}[{\bf z}_{k,t}{\bf z}_{k,t}^H]$  \cite{koldovsky2019TSP}. ${\bf C}_{\bf z}^{k,t}$ is an unknown nuisance parameter. The model densities are put into the likelihood function which is derived based on the fact that ${\bf s}_t$ and ${\bf z}_{k,t}$ are independent, by which the contrast function is obtained in the form
\begin{multline}\label{eq:contastCSV}
    \mathcal{C}\left(\{{\bf w}_k,{\bf a}_{k,t}\}_{k,t}\right) =\Bigg<\hat{\rm E}\left[\log f\left(\left\{\frac{\hat{s}_{k,t}}{\hat\sigma_{k,t}}\right\}_k\right)\right]  -\sum_{k=1}^{K}\log\hat\sigma_{k,t}^2\\ -\sum_{k=1}^{K} \hat{\rm E}\left[\hat{\bf z}_{k,t}^H({\bf C}_{\bf z}^{k,t})^{-1}\hat{\bf z}_{k,t}\right]  
    + (d-2)\sum_{k=1}^{K}\log |\gamma_{k,t}|^2\Bigg>_t + C,
\end{multline}
where $\hat{\cdot}$ denotes the estimate of the argument; $C$ is a constant independent of the parameter vectors.

The estimates of ${\bf a}_{k,t}$ and ${\bf w}_k$ are sought through finding optimum points of \eqref{eq:contastCSV}. However, since there are many spurious solutions, it is often necessary to impose the orthogonal constraint (OGC). It enforces that $\hat{\rm E}[\hat{s}_{k,t}\hat{\bf z}_{k,t}]=0$ for all $k$ and $t$. The OGC involves the distortionless constraint and makes ${\bf a}_{k,t}$ fully dependent on ${\bf w}_k$ through
\begin{equation}\label{eq:couplingaCSV}
	{\bf a}_{k,t}=\frac{\wCxkt{\bf w}_k}{{\bf w}_k^H\wCxkt{\bf w}_k},
\end{equation}
or vice verse \cite{koldovsky2019TSP}. For more details, see Section~III in \cite{koldovsky2020fastdiva}.

\subsection{QuickIVE}
The key approach in the algorithm development is that the gradient and Hessian matrices of \eqref{eq:contastCSV} w.r.t. ${\bf w}_k$ are computed without imposing the OGC. That is, ${\bf a}_{k,t}$ are treated as independent variables; only the link $\gamma^{k,t}=(1-({\bf h}^{k})^H{\bf g}^{k,t})/(\beta^{k})^*$ is considered due to the distortionless constraint.
In the following, detailed computations are skipped due to the limited space; they will appear in a forthcoming publication.

After some computations, we obtain 
\begin{equation}
\frac{\partial\mathcal{C}}{\partial{\bf w}_{k}^H} = 
\Big<\Re\{\hat\nu_{k,t}\}{\bf a}_{k,t}-\hat{\rm E}\left[\phi_k\left(\left\{\bar{s}_{k,t}\right\}_k\right) \bar{\bf x}_{k,t}\right]\Big>_t,\label{eq:gradient2}
\end{equation}
where $\bar{s}_{k,t}=\hat{s}_{k,t}/\hat\sigma_{k,t}$, $\bar{\bf x}_{k,t}={\bf x}_{k,t}/\hat\sigma_{k,t}$,  $\phi_k(\{s_{k}\}_k)=-\frac{\partial}{\partial s_k}\log f(\{s_{k}\}_k)$ is the score function; $s_k$ stands for the $k$th complex argument of $f(\cdot)$, and 
\begin{equation}\label{eq:nu}
\hat\nu_{k,t}=\hat{\rm E}[\phi_k(\{\bar{s}_{k,t}\}_k)\bar{s}_{k,t}].    
\end{equation}
It is necessary that when $N_b\rightarrow+\infty$ and when ${\bf a}_{k,t}$ and ${\bf w}_k$ correspond to their true values, the gradient is zero. This is satisfied when $\nu_{k,t}=1$. We therefore consider a substitution $\phi_{k}\leftarrow\hat\nu_{k,t}^{-1}\phi_{k}$, which can be justified by a more specific choice of the model density $f(\cdot)$ (in fact, $f(\cdot)$ need not be specified as it is not used explicitly). After this change, the gradient reads
\begin{equation}
\nabla_k= 
\Big<{\bf a}_{k,t}-\hat\nu_{k,t}^{-1}\hat{\rm E}\left[\phi_k\left(\left\{\bar{s}_{k,t}\right\}_k\right) \bar{\bf x}_{k,t}\right]\Big>_t.\label{eq:gradient2mod}
\end{equation}

The Hessian matrices $\frac{\partial \nabla_k^H}{\partial {\bf w}_k}$ and $\frac{\partial \nabla_k^T}{\partial {\bf w}_k}$  are computed under these additional rules: (a) $\hat\nu_{k,t}$ in \eqref{eq:gradient2mod} is treated as a constant, and (b) the analytic forms of the Hessian matrices when $N_b\rightarrow+\infty$ and ${\bf a}_{k,t}$ and ${\bf w}_k$ correspond to their true values are considered. Then,
\begin{equation}
{\bf H}^k_2=\frac{\partial \nabla_k^T}{\partial {\bf w}_k}=
\Big<-\nu_{k,t}^{-1}\frac{\partial}{\partial {\bf w}_k}{\rm E}\left[\phi_k\left(\left\{\bar{s}_{k,t}\right\}_k\right) \bar{\bf x}_{k,t}^T\right]\Big>_t.\label{eq:hessian2} 
\end{equation}
Using Appendix~A in \cite{koldovsky2020fastdiva}, it follows that 
\begin{equation}\label{eq:hessian3} 
    \frac{\partial}{\partial {\bf w}_k}{\rm E}\left[\phi_k\left(\left\{\bar{s}_{k,t}\right\}_k\right) \bar{\bf x}_{k,t}^T\right]=\frac{\rho_{k,t}}{\sigma^2_{k,t}}({\bf C}^{k,t})^*+\zeta_{k,t}{\bf a}_{k,t}^*{\bf a}_{k,t}^T,
\end{equation} 
where
\begin{equation}\label{eq:rho}
    \rho_{k,t} = {\rm E}\left[\frac{\partial\phi_k(\{\bar{s}_{k,t}\}_k)}{\partial s_k^*} \right].
\end{equation}
Similarly, it can be shown that the second Hessian matrix ${\bf H}^k_1=\frac{\partial \nabla_k^H}{\partial {\bf w}_k}=\left<\eta_{k,t}{\bf a}_{k,t}^*{\bf a}_{k,t}^T\right>_t$; $\zeta_{k,t}$ and $\eta_{k,t}$ are constants depending on the distribution of $s_{k,t}$ and on $\phi_k$. 

We now simplify the algorithm derivation through ignoring the rank-1 terms in ${\bf H}^k_1$ and in \eqref{eq:hessian3}. Hence, we take ${\bf H}^k_1=0$ and ${\bf H}^k_2=-\left<\frac{\rho_{k,t}}{\nu_{k,t}\sigma^2_{k,t}}({\bf C}^{k,t})^*\right>_t$. By \cite{li2008}, the Newton-Raphson direction is then given by $-(({\bf H}^k_2)^*)^{-1}\nabla_k$. Using the sample-based estimates instead of the corresponding expectation values, the update rule for the separating vector ${\bf w}_k$ therefore reads
\begin{equation}\label{eq:newtonupdate}
{\bf w}_k\leftarrow {\bf w}_k + \left<\frac{\hat\rho_{k,t}}{\hat\nu_{k,t}^*}\frac{\widehat{\bf C}^{k,t}}{\hat\sigma^2_{k,t}}\right>_t^{-1}\nabla_k.
\end{equation}

To summarize, the complete update rule in the proposed algorithm proceeds in the following steps. Starting from the given values of ${\bf w}^k$, for every $k=1,\dots, K$,
\begin{enumerate}
    \item update ${\bf w}^k$ according to \eqref{eq:newtonupdate},
    \item to fix the scaling ambiguity, normalize the separating vector as ${\bf w}_k \leftarrow {\bf w}_k/\sqrt{T\left<\sigma^2_{k,t}\right>_t},$
    \item and update the mixing vectors using \eqref{eq:couplingaCSV},
\end{enumerate}
until convergence. The stopping criterion as in \cite{koldovsky2020fastdiva} is used.


\section{QuickIVE versus Fixed-Point Algorithms}\label{sec:comparison with fastdiva}
FastDIVA (Fast Dynamic IVA) is a fixed-point algorithm that has been recently derived in \cite{koldovsky2020fastdiva}. For $T=1$ and $K=1$, FastDIVA coincides with the famous FastICA \cite{hyvarinen1999}, in the real-valued case, and is closely related to the complex-valued FastICA \cite{bingham2000} and FastIVA \cite{lee2007fast} in the complex-valued case and when $K>1$, respectively. Its one-unit variant, which performs the extraction of one SOI, updates the separating vectors according to
\begin{equation}\label{eq:updatefinal}
    {\bf w}_k \leftarrow {\bf w}_k - \left<\left(\frac{\hat\nu_{k,t}-\hat\rho_{k,t}}{\hat\nu_{k,t}}\right)^*\frac{\widehat{\bf C}_{k,t}}{\hat\sigma_{k,t}^2}\right>_t^{-1}\nabla^k.
\end{equation}

QuickIVE and One-unit FastDIVA are optimizing the same contrast function given by \eqref{eq:contastCSV}. Provided that both algorithms converge, they must therefore achieve the same asymptotic accuracy in terms of the mean interference-to-signal ratio, as derived in Section~IV in \cite{koldovsky2020fastdiva} and confirmed by the experiments therein. The algorithms differ in the speed of convergence because of the different optimization approaches: In FastDIVA, the gradient and Hessians of \eqref{eq:contastCSV} are computed when the OGC \eqref{eq:couplingaCSV} is imposed while, in QuickIVE, only the weaker distortionless constraint is considered\footnote{Interestingly, both computations yield the same gradient but different Hessians; cf. \eqref{eq:gradient2mod} in the present paper and (43) in \cite{koldovsky2020fastdiva}.}. It means that the mixing and separating vectors are updated simultaneously in FastDIVA, which might explain its faster convergence compared to QuickIVE.

For further comparison, let us consider the real-valued case when $T=1$ and $K=1$, which corresponds to the fundamental problem of extraction of one independent component (ICE). Here, the indices $k$ and $t$ and the conjugate operator can be omitted, and, owing to the normalization of the output signal scale, $\hat\sigma^2=1$. Then, the QuickIVE update rule \eqref{eq:newtonupdate} simplifies to 
\begin{equation}\label{eq:newtonupdatesimple}
{\bf w}\leftarrow {\bf w} + \frac{\hat\nu}{\hat\rho}\widehat{\bf C}^{-1}\nabla,
\end{equation}
and the FastDIVA (FastICA) update \eqref{eq:updatefinal} simplifies to 
\begin{equation}\label{eq:updatefinalsimple}
    {\bf w} \leftarrow {\bf w} + \frac{\hat\nu}{\hat\rho-\hat\nu}\widehat{\bf C}^{-1}\nabla.
\end{equation}
An important difference of FastDIVA compared to QuickIVE (and also compared to other algorithms) stems from the numerator value $\hat\rho-\hat\nu$. It can be shown that, for $N\rightarrow+\infty$, this value is zero when the SOI is Gaussian \cite{tichavsky2006,rati2007}. It means that, if the distribution of the SOI becomes Gaussian for a moment (e.g. in an on-line processing regime), FastDIVA tends to change the direction of the update abruptly, most likely, towards a different independent source. 


\section{Experiments}
\subsection{Convergence speed}
To compare algorithms in terms of the speed of convergence, $K=6$ complex-valued static ($T=1$) mixtures of $d=6$ signals of length $N=1000$ were generated in $100$ trials; the compared algorithms are One-unit FastDIVA \cite{koldovsky2020fastdiva} (FastDIVA), QuickIVE, OverIVA \cite{scheibler2019overiva} (equivalent to AuxIVE from \cite{jansky2020}), and BOGIVE$\_{\bf w}$ \cite{koldovsky2019icassp,jansky2020}. The algorithms were initialized in a vicinity of the desired SOIs and applied to extract them. The interference-to-signal ratio (ISR) has been evaluated in each iteration.
In each trial, all signals were generated from the Laplacean distribution, and the SOIs from different mixtures were mixed by a random unitary matrix to make them dependent (but uncorrelated). Then, the signals were mixed by random mixing matrices. In this comparison, the static mixtures are considered, i.e., $T=1$.

\begin{figure}
    \centering
    \includegraphics[width=0.9\linewidth]{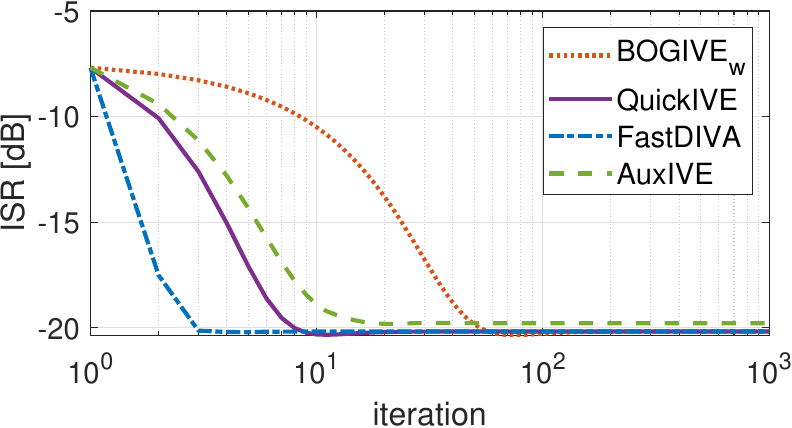}
    \caption{Averaged Interference-to-Signal Ratio as a function of the iteration index. The compared algorithms are initialized at the same point in the vicinity of the SOIs. The average is taken over $100$ independent trials. }
    \label{fig:fig1}
\end{figure}

Fig.~\ref{fig:fig1} shows the ISR as a function of iteration number; the ISR is averaged over trials and all SOIs in the $K=6$ mixtures. The results show typical speed of convergence: FastDIVA converges in very few iterations; QuickIVE and OverIVA converge in about 10 iterations; the slowest convergence shows the gradient-based BOGIVE$\_{\bf w}$. 

It is worth noting that FastDIVA, QuickIVE, and BOGIVE$\_{\bf w}$ finally achieve the same ISR. This is because they use same nonlinearity $\phi_k(\{\bar{s}_{k,t}\}_k)=\frac{\bar{s}_{k,t}^*}{1+\sum_{k=1}^K|\bar{s}_{k,t}|^2}$. AuxIVE is using a different nonlinearity, which meets the additional conditions for the auxiliary-function based optimization \cite{ono2011stable}. Therefore, the final ISR achieved by AuxIVE is different from that of the other algorithms.

\subsection{On-line source extraction}
\label{sub:online}
In this experiment, an on-line blind source extraction problem is simulated where the SOI is being extracted from a continuously varying real-valued mixture ($K=1$). In a trial, the mixture of dimension $d=5$ and length $N=128000$ is generated: The SOI is Laplacean up to an interval of length $16000$ samples beginning at $N/2$, where it is Gaussian. The other four signals are Laplacean. The mixing matrix at sample $n$ is given by ${\bf A}_n=[{\bf a}_n {\bf Q}_{\lceil 8*n/N\rceil}]$, where $\lceil x \rceil$ denotes the least integer greater than or equal to $x$,
\begin{align}
    {\bf a}_n &= {\bf a}_{n-1} + {\bf e}_n, \\
    {\bf Q}_i &= {\bf Q}_{i-1} + {\bf E}_i,
\end{align}
where the elements of ${\bf e}_n$ and ${\bf E}_n$ are drawn from $\mathcal{N}(0,\sqrt{0.005})$ and $\mathcal{N}(0,\sqrt{0.01})$; the initial ${\bf Q}_1$ and ${\bf a}_1$ are generated from $\mathcal{N}(0,1)$. Hence, the SOI performs a continuous Brownian motion while the background signals perform a step movement after every $N/8$ samples.

\begin{figure}
    \centering
    \includegraphics[width=\linewidth]{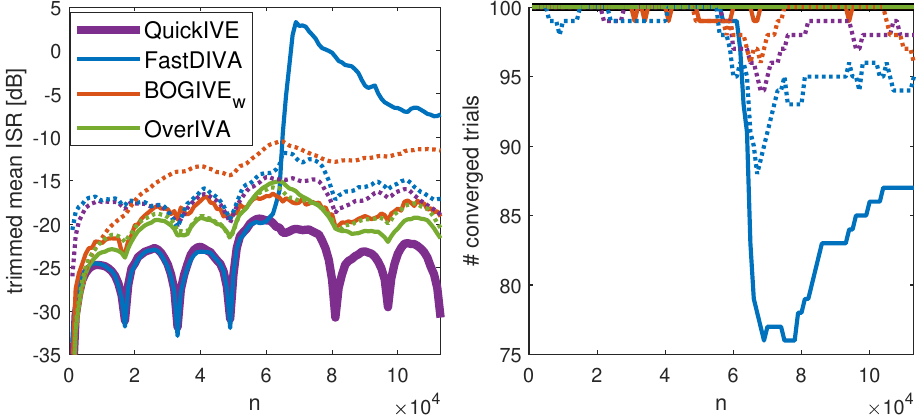}
    \caption{Trimmed mean of ISR and the number of converged trials as functions of $n=1,\dots,N$; $N=128000$; the SOI is Gaussian for $n=N/2,\dots,N/2+15999$. Solid and dotted lines correspond to algorithms' variants, respectively, exploiting CSV with $T=16$ and the static variants with $T=1$. For $T=16$, OverIVA is represented by Block AuxIVE \cite{jansky2020}. The significant local minima of ISR (e.g. by QuickIVE) when $n$ is an integer multiple of $16000$ correspond to situations when the batch coincides with an interval in which ${\bf Q}_i$ is constant (i.e. the background sources are static).}
    \label{fig:fig2}
\end{figure}

The compared algorithms are deployed sequentially batch-by-batch performing one iteration per batch; the batch size is $16000$; the shift is $1000$. The CSV model is used with $T=16$ (i.e. the batches are divided into blocks whose length corresponds with the shift length). The methods are initialized by a slightly perturbed true separating vector at time $n=1$. ISR achieved by the algorithms is evaluated per batch. Its 20\%-trimmed mean value computed over $100$ trials as a function of $n$ is shown in Fig.~\ref{fig:fig2} together with the number of converged trials: a trial is marked as converged it when ISR is smaller than $-3$~dB.  

The results reveal the sensitivity of FastDIVA to the Gaussianity of the SOI as its averaged ISR steeply increases at $n=N/2$ and the number of converged trials drops down. In fact, FastDIVA tends to extract a different Laplacean source since $n=N/2$ and keeps extracting it in most subsequent batches. The other methods show lower sensitivity to the Gaussianity of the SOI.

The results also demonstrate the benefit of using the CSV mixing model in on-line extraction. Significantly lower ISRs are achieved by the variants of the algorithms where the CSV mixing model is exploited with $T=16$ compared to when $T=1$. 

\subsection{On-line speech extraction in real environment}
We have realized a similar experiment with two speakers in an open-space $12$x$8$x$2.6$~m room with the reverberation time $\mathrm{T_{60}}\approx 500 \mathrm{ms}$. A dense microphone array consisting of 48 MEMS microphones arranged in an  $8\times 6$ vertical planar grid was used to record the speakers simulated by a loudspeaker positioned at $\sim1$~m away from the microphone array. The SOI, female speaker, was positioned at $-30^\circ$ from the front of the array. The female speech, which is $40$~s long, involved white Gaussian noise from 15~$s$ through $20$~s. This simulated an interval where the SOI is Gaussian as in the previous Section~\ref{sub:online}. An interfering male speaker (IR) was placed at $60^\circ$; its $40$~s long utterance was unaltered. Both sources were recorded separately at $48$~kHz sampling frequency, downsampled to $16$~kHz and mixed with initial SIR~$= 5$~dB. The processing was performed in the  short-time Fourier transform domain with a frame length of $512$ samples and $128$ samples overlap.

\begin{figure}
    \centering
    \includegraphics[width=1\columnwidth]{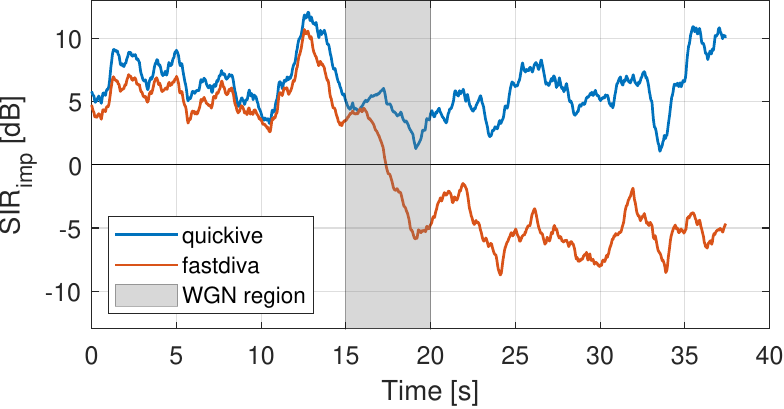}
    \caption{SIR improvement by QuickIVE and FastDIVA when the speech of the SOI is replaced by white Gaussian noise (WGN) for 5 seconds. }
    \label{fig:fig3}
\end{figure}

The SOI was extracted by QuickIVE and FastDIVA. Both algorithms used the same settings: $600$ frames per batch (corresponding to $4.8$~seconds of data context), $300$ frames shift, and $300$ frames for each CSV block ($T = 2$); each algorithm performed $3$ iterations per batch. The algorithms were initialized by an oracle MPDR beamformer. 
The results in terms of output sample-wise SIR improvement smoothed by a moving-average filter of a $2$-second length are shown in Fig.~\ref{fig:fig3}.

As expected, both algorithms do not adapt correctly when the SOI is Gaussian. FastDIVA exhibits a fast convergence speed, which 
causes that the method quickly converges to the IR when the SOI is Gaussian. Furthermore, despite the SOI being dominant in the mixture, FastDIVA keeps converging to the IR for the rest of the recording. QuickIVE converges slightly slower and does not exhibit as high of a sensitivity to the SOI Gaussianity as FastDIVA. This causes QuickIVE to continue the extraction of the SOI after the testing interval. 

\section{Conclusions}
We have proposed the QuickIVE algorithm for independent vector extraction allowing for the CSV mixing model, which is flexible for on-line processing of dynamic mixtures. QuickIVE provides a useful alternative to the fixed-point and auxiliary-function based methods. While its convergence speed is comparable or might be slightly slower, it shows robustness to a short-term Guassianity of the SOI. Compared to OverIVA, AuxIVA or Block AuxIVE, there are no additional requirements on the nonlinearity such as finding an appropriate majorizing function. 
In the experiment with a dense microphone array, the method shows an efficient performance and a reasonable computational burden because, compared to ICA/IVA methods, it extracts only the desired SOI.
\balance

\end{document}